\documentclass[fleqn,twoside]{article}
\usepackage{espcrc2}
\usepackage{amssymb}

\usepackage{psfig}
\usepackage[figuresright]{rotating}

\newcommand{\om}{\omega}
\newcommand{\be}{\begin{equation}}
\newcommand{\ee}{\end{equation}}
\newcommand{\bea}{\begin{eqnarray}}
\newcommand{\eea}{\end{eqnarray}}

\newcommand{\vecnul}{{\mathbf 0}}
\newcommand{\bra}{\langle}
\newcommand{\ket}{\rangle}
\newcommand{\hm}{\hspace{-0.6cm}}
\newcommand{\nn}{\nonumber}

\newcommand{\AmS}{{\protect\the\textfont2
  A\kern-.1667em\lower.5ex\hbox{M}\kern-.125emS}}

\title{Transport coefficients from the lattice?\thanks{Talk presented by 
G.~A.\ at Lattice 2002, June 24--29, 2002, MIT. Based on 
Ref.~\cite{Aarts}. }
}

\author{Gert Aarts and Jose M.\ Mart{\'\i}nez Resco\address{Department 
of Physics, The Ohio State University, 174 West 18th Avenue, Columbus, OH 
43210, USA}
}

\begin{document}

\begin{abstract}
The prospects of extracting transport coefficients from euclidean lattice 
simulations are discussed. Some general comments on the reconstruction 
of spectral functions using the Maximal Entropy Method are given as well.
\end{abstract}

\maketitle

1.\ 
In field theory transport coefficients are proportional to the slope 
of appropriate spectral functions at zero frequency and zero spatial 
momentum (Kubo relation). Examples are the electrical conductivity,
\be
 \label{eqcond}
 \sigma = \frac{1}{6}\frac{\partial}{\partial \om} 
 \rho_{\rm em}(\om,\vecnul)\Big|_{\om=0},
\ee
where $\rho_{\rm em}(x,y) = \bra [j_{\rm em}^i(x), j_{\rm em}^i(y)]\ket$
is the spectral function associated with the 
electromagnetic current $j_{\rm em}^i = \bar\psi \gamma^i\psi$,
and the shear viscosity,
\be
 \eta = \frac{1}{20}\frac{\partial}{\partial \om}
 \rho_{\pi\pi}(\om,\vecnul)\Big|_{\om=0},
\ee
where $\rho_{\pi\pi}(x,y) = \bra [\pi_{kl}(x), \pi_{kl}(y)]\ket$ with
$\pi_{kl}$ the traceless part of the spatial energy-momentum tensor.

The euclidean two-point function and the spectral function (both at zero 
momentum) are related via a dispersion relation and
\be
 \label{eqdisp}
 G(\tau) = \int_0^\infty \frac{d\om}{2\pi}\, K(\tau,\om)\rho(\om),
\ee
with the kernel
\be
 \label{eqkernel}
 K(\tau, \om) = e^{\om \tau}n_B(\om) + e^{-\om\tau}\left[1+n_B(\om)\right],
\ee
where $n_{B}(\om)= 1/[e^{\om/T}- 1]$ is the Bose distribution. The first
attempt to compute transport coefficients on the lattice using Eq.\
(\ref{eqdisp}) was made some time ago by Karsch and Wyld
\cite{Karsch:1986cq}, by fitting a three-parameter ansatz for the spectral
function to the euclidean lattice data. This approach was pursued more
recently in Ref.\ \cite{Nakamura:1996na}. A modern way to attack this
problem would of course be to use the Maximal Entropy Method
\cite{Asakawa:2000tr}. Once the spectral function is reconstructed for all
$\om$, the transport coefficient is in principle determined.

Two obvious questions are: What is the spectral function expected to look
like at high temperature? How does the transport coefficient, or in
general the low-frequency part of the spectral function, show up in the 
euclidean correlator?

\begin{figure*}[t]
\centerline{\psfig{file=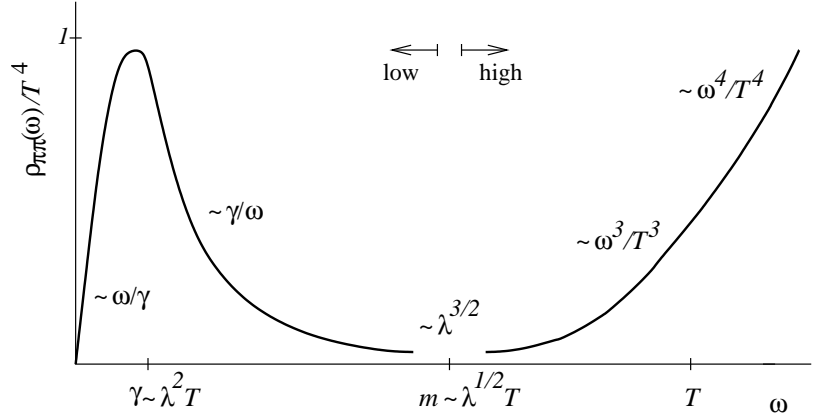,height=5.5cm}}
\vspace{-0.5cm}
\caption{Sketch of the spectral function $\rho_{\pi\pi}(\om)/T^4$ versus 
$\om$ in a weakly-coupled scalar $\lambda\phi^4$ theory 
(not to scale). 
The shear viscosity is given by the slope at zero frequency. 
See the main text for discussion.
}
\label{figrhoxfig}
\end{figure*}

2.\
To answer the first question, we calculated the spectral function relevant 
for the shear viscosity at high temperature in weakly-coupled scalar and 
nonabelian theories \cite{Aarts}. The results for the scalar theory are 
sketched in Fig.~\ref{figrhoxfig}. The contribution at higher frequencies, 
i.e.\ $\om\gtrsim 2m$ where $m$ is the thermal mass, arises from (inverse) 
decay processes. At large enough $\om$ the spectral function increases as 
$\rho_{\pi\pi}(\om) = (\om^4/96\pi) \left[ 1+2n_B(\om/2) \right]$. 
The origin of the contribution at lower frequencies is scattering of 
fast-moving particles with soft bosons in the plasma. The 
approximate shape of the spectral function in this region is given by
\be
 \label{eqpinch}
 \frac{\rho_{\pi\pi}(\om)}{T^4} \sim \frac{\om\gamma}{\om^2+4\gamma^2}
 \;\;\;\;\;\;\;\;\;\;\;\;\;\;\;(\om \ll T),
\ee
where $\gamma$ is the thermal damping rate.
The particular form in Eq.\ (\ref{eqpinch}) is due to the presence of 
poles in the complex energy-plane that pinch the real axis from above and 
below.
The denominator indicates the distance between these so-called 
pinching poles.
The viscosity is determined by the slope at zero frequency and is 
proportional to $1/\gamma$. 
A quantitative analytical calculation in this regime is 
complicated since due to the pinching poles the loop expansion 
breaks down and all ladder diagrams with uncrossed rungs contribute at 
leading order in the coupling \cite{Jeon:if}.
In the scalar theory, the low- and the high-frequency contribution match 
parametrically around the threshold for decay, $\om=2m$. 
A general ansatz describing the low-frequency contribution is
\be
 \label{eqansatz}
 \frac{\rho^{\rm low}_{\pi\pi}(x)}{T^4} =
 x\frac{b_1+b_2x^2+b_3x^4+\ldots}{1+c_1x^2+c_2x^4+c_3x^6+\ldots},
\ee
where $x=\om/T$ and $b_i=c_i=0$, $i> n$ for given~$n$. 
For gauge theories the overall characteristic shape remains as in Fig.\ 
\ref{figrhoxfig} but the dependence on the coupling constant 
$g^2\sim\lambda$ differs somewhat~\cite{Aarts}.

\begin{figure}[t]
\centerline{\psfig{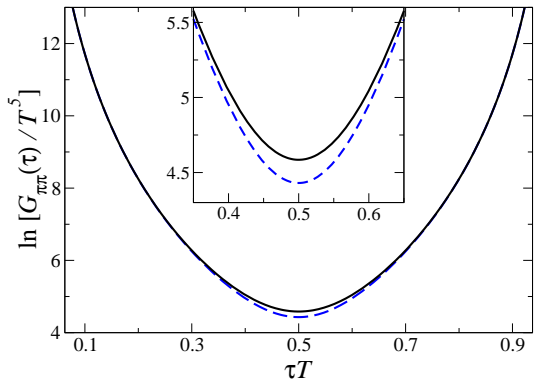}}
\vspace{-0.5cm}
\caption{
Logarithm of $G_{\pi\pi}(\tau)/T^5$ versus $\tau T$
in weakly-coupled $SU(3)$ (Eq.\ (\ref{eqG}), full line). The dashed 
line is the result without the constant contribution 
originating from the low-frequency part of the spectral function. The 
inset shows a blowup around $\tau T=0.5$.
}
\label{figGE}
\end{figure}

3.\
We now address the second question raised in the beginning. At leading 
order in the coupling constant the euclidean correlator can be computed 
exactly using Eq.~(\ref{eqdisp}), and we find
\bea
 \nn G_{\pi\pi}(\tau) =  &&\hm 
 \frac{a_1\pi^2T^5}{3\sin^5 u}
 \Big\{ (\pi-u)\left[11\cos u + \cos 3u\right] \\
 &&\hm +\, 6\sin u+ 2\sin 3u \Big\} + \frac{4a_2\pi^2}{45}T^5,
 \label{eqG}
\eea
with $u=2\pi \tau T$. For the scalar theory $a_1 = a_2 = 1$, while 
for an $SU(N_c$) theory $a_1 = 6a_2 = 12(N_c^2-1)$. 
The $\tau$-dependent terms arise entirely from the high-frequency part of 
the spectral function. The $\tau$-independent term originates from the 
low-frequency part. The fact that the low-frequency part of the 
spectral function leads to a constant term in the euclidean 
correlator is easily understood.
Since for small frequencies $\om\ll T$ the kernel (\ref{eqkernel}) can be 
expanded as
\be
 \label{eqexp}
 K(\tau,\om) = \frac{2T}{\om} 
 + {\cal O}\left(\frac{\om}{T}\right),
\ee
and the dominating term at small $\om$ is $\tau$-independent,
the low-frequency part of the spectral function corresponds in the 
euclidean correlator to a constant term proportional to $\int d\om\,
\rho_{\pi\pi}(\om)/\om$. This particular constant cannot be easily 
disentangled from the high-frequency contribution.
This is illustrated in Fig.~\ref{figGE}, where
we plot the analytical result for $G_{\pi\pi}(\tau)$ in the case of
$SU(3)$. The tiny difference between the full and the dashed lines is due
to the constant term in Eq.~(\ref{eqG}). Once again, this term originates
from the low-frequency part of the spectral function and reflects $\int
d\om\, \rho_{\pi\pi}(\om)/\om$, not $\rho_{\pi\pi}(\om)$ itself. 
Therefore, we find that $G_{\pi\pi}(\tau)$ is remarkably insensitive 
to details of $\rho_{\pi\pi}(\om)$ when $\om\ll T$ and we conclude that 
it is extremely difficult to extract transport coefficients in 
weakly-coupled field theories from the euclidean lattice.

4.\
The results described above are generic and not specific for the
correlators we considered. The fact that the low-frequency part of a
spectral function corresponds to a constant term in the euclidean
correlator relies solely on the expansion (\ref{eqexp}). In our opinion
this result is a potential problem for the Maximal Entropy Method when the
reconstruction of low-frequency parts of spectral functions is attempted.
We emphasize that the difficulty is not a numerical one, e.g.\ due to a
finite number of lattice points in the imaginary-time direction.
Furthermore, it is easy to see that at finite temperature the presence of 
pinching poles in correlators of composite operators that are bilinear in 
the fields is quite common.  
From general considerations it follows that the effect of pinching poles 
is a low-frequency contribution as in Eqs.\ (\ref{eqpinch}, 
\ref{eqansatz}) and Fig.~\ref{figrhoxfig}.

Consider for instance the electromagnetic current-current correlator 
(or any two-point function of fermion bilinears) in the deconfined 
quark-gluon plasma. For large frequencies a perturbative calculation gives 
$\rho_{\rm em}(\om) \sim \om^2\left[1-2n_F(\om/2)\right]$,
where $n_F(\om) = 1/[e^{\om/T}+1]$ is the Fermi distribution 
and the proportionality factor depends on the number of fermions, 
colours, etc. For this reason it is customary to present this kind of 
spectral functions as $\rho_{\rm em}(\om)/\om^2$. However, due to pinching 
poles also this spectral function is expected to have a structure at small 
frequencies as in Eqs.\ (\ref{eqpinch}, \ref{eqansatz}) and 
Fig.~\ref{figrhoxfig}, where in this case $\gamma$ is the fermion damping 
rate. For very small frequencies the spectral function can be expanded as  
\be
 \label{eqsmall} 
 \frac{\rho_{\rm em}(\om)}{T^2} = d_1 \frac{\om}{T} + 
 d_3\left(\frac{\om}{T}\right)^3 + 
 d_5\left(\frac{\om}{T}\right)^5 + 
 \ldots,
\ee
where $d_1$ is proportional to the electrical conductivity. The
coefficients $d_i$ are nonzero at finite temperature only and represent
repeated scattering of the fast-moving on-shell fermions with soft gauge
bosons in the deconfined phase of the high-temperature plasma. Because of
this behaviour, $\rho_{\rm em}(\om)/\om^2$ diverges as $1/\om$ when
$\om\to 0$. Singular behaviour at very small frequencies of spectral 
functions of fermion bilinears that are normalized with $1/\om^2$
has indeed been found \cite{Asakawa:2002xj} although the statistical 
significance of these results is still uncertain.

\section*{Acknowledgements}

We thank Guy Moore for discussions. G.~A.\ is supported by the Ohio State
University through a Postdoctoral Fellowship and J.~M.~M.~R. is supported
by a Postdoctoral Fellowship from the Basque Government.

\end{document}